\documentclass[letter,11pt]{article}
\pdfoutput=1
\usepackage{heppub} 
\allowdisplaybreaks
\usepackage{hyperref}
\usepackage{graphicx}
\usepackage{amsmath}
\usepackage{amssymb} 
\usepackage[normalem]{ulem}
\usepackage{slashed}
\usepackage[utf8]{inputenc}
\usepackage[T1]{fontenc}
\usepackage{subcaption}
\usepackage{scalerel,stackengine}
\usepackage{multirow}
\usepackage{array}
\usepackage{orcidlink}

\newcommand{\nn}{\nonumber}

\def\bea#1\eea{\begin{align}#1\end{align}}
\newcommand{\bs}{\boldsymbol}

\preprint{IQuS@UW-21-133}

\title{Quantum Simulation of QCD in Axial Gauge}

\author{Xiaojun Yao \orcidlink{0000-0002-8377-2203}}

\affiliation{InQubator for Quantum Simulation, Department of Physics, University of Washington, Seattle, WA 98195, USA}

\emailAdd{xjyao@uw.edu}

\abstract{We study quantum simulation of SU(3) non-Abelian gauge theory dynamically coupled with fundamental fermions in $3+1$ dimensions by employing the lattice Hamiltonian in axial gauge under fixed boundary conditions, which avoids Gauss's law constraints. The temporal component of the gauge field is analytically solved in terms of independent field degrees of freedom and a lattice regulated Green's function. The axial gauge condition is trivially maintained in time evolution, even under Trotterization. The gauge field degrees of freedom are expressed in the local field basis and can be efficiently transformed into the canonical conjugate momentum basis by local quantum Fourier transforms. We prove the number of qubits needed for describing all states up to an energy $E$ with an accuracy $\epsilon$ on a lattice of volume $V$ at bare coupling $g$ is bounded as $16n_A V + 12n_f V$, where $n_A \approx \log_2 (\frac{64 E' V^{4/3}}{\pi^2\epsilon} + \frac{32\sqrt{2}g n_f E'^{1/2} V^{7/6}}{\sqrt{3}\pi^3\epsilon^{1/2}} ) $ is the number of qubits needed for each independent gauge field per site with a shifted energy $E'$, and $n_f$ denotes the number of fermion flavors. We then analyze a quantum algorithm for time evolution that is based on Trotterization, quantum Fourier transform, and Jordan-Wigner transformation, for which quantum circuits can be explicitly constructed under arbitrary gauge field truncation and digitization. We find the numbers of CNOT and single-qubit rotation gates scale as $O(n_A^4 V^{4/3}) + O(V^{5/3})$ and $O(n_A^4 V^{4/3})$, respectively, per Trotter step for fixed $n_f\leq 6$. Together with the bounded Trotterization error, we conclude that quantum resources needed for simulating real-time dynamics of lattice QCD scale polynomially with volume, energy, time, accuracy, and bare Hamiltonian parameters.
}

\begin{document}
\maketitle

\section{Introduction}
\label{sec:intro}
Real-time Hamiltonian lattice simulation is becoming a useful nonperturbative tool to study field theories~\cite{Zohar:2015hwa,Banuls:2019bmf,Klco:2021lap,Bauer:2022hpo,Bauer:2023qgm,DiMeglio:2023nsa,Beck:2023xhh}, in particular for non-equilibrium dynamics~\cite{Halimeh:2025vvp} that generates simultaneous high entanglement and high nonstabilizerness~\cite{Ebner:2025pdm,Grieninger:2026bdq,Robin:2026lqp}. Much progress has been made in $1+1$ dimensions by using either classical methods such as tensor networks~\cite{Banuls:2013jaa,Buyens:2013yza,Kuhn:2015zqa,Pichler:2015yqa,Magnifico:2019kyj,Honda:2021aum,Dempsey:2022nys,Dempsey:2023gib,Jha:2024jan,Florio:2025hoc,Banuls:2025wiq,Abel:2025pxa,Grieninger:2025mbm,Gupta:2026tcg,Fujikura:2026tqk,Yao:2026yld} or quantum computing~\cite{Farrell:2024fit,Zemlevskiy:2024vxt,Illa:2024kmf,Farrell:2024mgu,Hayata:2024smx,Farrell:2025nkx,Chen:2025zeh,Abel:2025zxb,Li:2025sgo,Davoudi:2025rdv,Huie:2025yzn,Chai:2025kbi,Chen:2026tvd,Andrade:2026kbz,Gonzalez:2026loy,Hayata:2026tly}. Attempts to simulate higher-dimensional field theories have started~\cite{Osborne:2022jxq,Ciavarella:2023mfc,Itou:2024psm,Mueller:2024mmk,Maiti:2024jwk,Ale:2024uxf,Turro:2024pxu,Spagnoli:2024mib,Budde:2024rql,Ebner:2024qtu,Yang:2025edn,DiMarcantonio:2025cmf,Illa:2025njz,NSSrivatsa:2025jhh,Balaji:2025afl,Balaji:2025yua,Ale:2025sxz,Yao:2025cxs,Cobos:2025krn,Horn:2026bfs,Orlando:2026ten,Cao:2026qky,Majcen:2026sfm,Joshi:2026hfe,Xu:2026ibi,Lewis:2026ync,Rouxinol:2026vjl,Rouxinol:2026eur,Turco:2026cte} with simulating Quantum Chromodynamics (QCD) as one of the ultimate goals, which is an SU(3) non-Abelian gauge theory coupled with a few Dirac fermions carrying fundamental colors in $3+1$ dimensions.

A widely studied setup for real-time simulation of lattice gauge theories is the Kogut-Susskind (KS) Hamiltonian~\cite{Kogut:1974ag}, which uses the temporal gauge Hamiltonian and the electric basis. As a consequence of the temporal gauge choice, Gauss's law at each vertex must be imposed, either by constraining the Hilbert space to contain only gauge invariant states, or maintaining all Gauss's law constraints in the Hamiltonian evolution that is usually Trotterized. In the electric basis, the gauge degree of freedom on each link is described in the basis of irreducible representations (irreps) and then plaquette operators can be written as raising and lowering operators for qubits with Clebsch-Gordan (CG) coefficients accounted for~\cite{Byrnes:2005qx}. The CG coefficients are quite numerous and need hard coding into quantum circuits. On trivalent lattices, the Gauss's law constraints can be completely ``integrated out'' for SU(2) and SU(3) truncated below the adjoint irrep~\cite{Klco:2019evd,Ciavarella:2021nmj,ARahman:2021ktn,ARahman:2022tkr,Muller:2023nnk,Kavaki:2024ijd,Illa:2025dou,Chen:2026hnh}, which at the same time reduces the number of qubits needed locally under a given truncation of the irreps. The price is in the implementation of the plaquette operator, which involves Wigner $6j$ symbols [and analogs for SU(3)] that turn into complicated controls and deep circuits in quantum computing. To overcome these difficulties in the original KS Hamiltonian, many approaches have been proposed such as the loop-string-hadron formulation~\cite{Raychowdhury:2019iki,Kadam:2022ipf,Kadam:2024ifg,Burbano:2024uvn} that expresses the Hamiltonian directly in terms of operators obeying Gauss's law, a mixed basis approach~\cite{Grabowska:2024emw,DAndrea:2023qnr,Froland:2025bqf} that is built on further fixing the spatial gauge redundancies (called maximal tree gauge~\cite{Creutz:1976ch}), using large $N_c$ expansion~\cite{Ciavarella:2024fzw,Ciavarella:2025bsg,Modi:2026syn}, $q$-deformed algebras~\cite{Zache:2023dko,Hayata:2023bgh,Hayata:2024fnh,Hayata:2026rmv}, finite subgroups of SU($N_c$)~\cite{Gustafson:2023kvd,Lamm:2024jnl,Perez:2025cxl}, variationally optimizing coupling-dependent bases~\cite{Haase:2020kaj,Fontana:2024rux,Miranda-Riaza:2025fus}, and the orbifold formalism~\cite{Bergner:2024qjl,Halimeh:2024bth,Bergner:2026duh}. In the last approach, quantum circuits for Trotterized time evolution can be analytically constructed in a straightforward way under an arbitrary truncation of the local gauge degrees of freedom.

From a different perspective, it is important to bound the number of qubits and gates for simulating certain physics processes at a given accuracy. This type of bound was first proved for $\lambda\phi^4$ scalar field theory~\cite{Jordan:2011ci,Jordan:2012xnu}, which later led to the proof that estimating the vacuum-to-vacuum transition amplitude in massive $\lambda\phi^4$ scalar field theory is BQP-complete~\cite{Jordan:2017lea}. More recently, a bound for lattice Quantum Electrodynamics (QED) was proved by using the Coulomb gauge~\cite{Yao:2025uxz}, which completely avoids the difficulty of imposing Gauss's law at each vertex mentioned above. However, applying the Coulomb gauge QCD Hamiltonian faces two significant challenges: inverting the covariant Laplacian (defined as the square of the covariant derivative) and the Gribov ambiguity~\cite{Gribov:1977wm,Vandersickel:2012tz}. 

Here we study quantum simulation of QCD by choosing the axial gauge $A_3^a=0$ (the superscript $a$ denotes an adjoint color). Axial gauge has been used before in $1+1$ dimensional simulations~\cite{Farrell:2022wyt,Farrell:2023fgd}, which is shown equivalent to the temporal gauge after Gauss's law has been integrated out~\cite{Lin:2024eiz}, which is easy to do in $1+1$ dimensions. The advantage of axial gauge becomes more significant in $2+1$ or higher dimensions as gauge fields become dynamical, which we pursue here. The advantage is simultaneously two-fold: First, the $A_0^a$ gauge field can be analytically expressed in terms of independent dynamical variables at each time, thus avoiding Gauss's law. Then we can use field (like position) basis for the independent gauge field variables $A_1^a$ and $A_2^a$, which can be efficiently transformed into the canonical conjugate variable (like momentum) basis by local quantum Fourier transforms. Construction of quantum gates for Trotterized time evolution is straightforward and can be analytically written out explicitly for arbitrary truncation and digitization of the field basis. Second, the setup enables us to prove bounds for the number of qubits and gates to describe all states up to a given energy and accuracy on a lattice, which have not been proved before in the same rigorous way as done for $\lambda \phi^4$ scalar field theory and QED. The main results here are Eq.~\eqref{eqn:nA_final}, Table~\ref{tab:costs}, and the quantum algorithm for simulating the time evolution of QCD.

This paper is organized as follows: In Sec.~\ref{sec:axial}, a brief review of the axial gauge QCD Hamiltonian in the continuum will be given, followed by a discussion of a Green's function contained in the Hamiltonian. Then in Sec.~\ref{sec:dd}, we will introduce a discretized lattice Hamiltonian containing a well-defined Green's function regulated on the lattice. We will also discuss truncation and digitization in the gauge field basis and how to map gauge and fermion fields onto qubits. Bounds on the truncation value and digitization step size for gauge fields will be proved in Sec.~\ref{sec:qubit_cost} for describing all states up to an energy and accuracy, which together give a bound on the number of qubits needed. Section~\ref{sec:time_and_gate} will analyze quantum circuits for implementing Trotterized time evolution and provide bounds on the number of CNOT and single-qubit rotation gates. Finally, we will conclude in Sec.~\ref{sec:conclusions} and discuss future prospects.

\section{QCD Hamiltonian in Axial Gauge}
\label{sec:axial}
Here we first briefly review the Hamiltonian of QCD in axial gauge, for which more details can be found in Ref.~\cite{Weinberg:1996kr}. Then we discuss a Green's function that appears in the axial gauge Hamiltonian. We start with the Lagrangian density of the theory in $3+1$ dimensions
\begin{align}
    \mathcal{L} = -\frac{1}{4}F^{\mu\nu a} F_{\mu\nu}^a + \sum_f \bar{\psi}_f (i \gamma^\mu D_\mu - m_f) \psi_f \,,
\end{align}
where the gauge field strength tensor is $F_{\mu\nu}^a = \partial_\mu A_\nu^a - \partial_\nu A_\mu^a + gf^{abc} A_\mu^b A_\nu^c$ with the $f^{abc}$ being the structure constants of the Lie algebra of the SU(3) gauge group. The Greek letters label the Lorentz indices, the lowercase letters in the superscripts indicate the adjoint color indices, and the subscripts $f$ in the fermion fields $\psi_f$ label the flavors that are summed over. Repeated indices are summed over. The covariant derivative is $D_\mu = \partial_\mu - ig A_\mu^a T^a$ with $T^a$ being the generator of the SU(3) group in the fundamental representation and $g$ being the gauge coupling.

The gauge theory is a constrained quantum system. The primary constraint is
\begin{align}
    \Pi_0^a \equiv \frac{\delta \mathcal{L}}{\delta(\partial_0A_0^a)} = 0 \,,
\end{align}
while the secondary constraint originates from the consistency requirement that $\Pi_0^a$ remains vanishing throughout time evolution, i.e., $\dot{\Pi}_0^a = \partial_0 \Pi_0^a = 0$, which can be obtained from the Euler-Lagrange equation associated with $A_0^a$ and gives
\begin{align}
    \partial_i \Pi_i^a + gf^{abc} A_i^b \Pi_i^c + gJ^{0a} = 0\,,
\end{align}
where we have introduced the conjugate momentum operator associated with $A_i^a$ and the fermion color current as\footnote{Some may prefer the Lorentz covariant notation $\Pi^{ia}$. It is just a definition and the whole Hamiltonian formalism does not look Lorentz invariant in any way.}
\begin{align}
    \Pi_i^a &\equiv \frac{\delta \mathcal{L}}{\delta(\partial_0A_i^a)} = -F^{0i a} = F_{0i}^a \,,\\
    J^{\mu a} &\equiv \sum_f \bar{\psi}_f \gamma^\mu T^a \psi_f \,,
\end{align}
respectively. The primary and secondary constraints commute so they are first-class constraints.

To quantize this constrained quantum system, we choose the axial gauge
\begin{align}
    A_3^a = 0 \,.
\end{align}
Then the independent dynamical variables for the gauge field are $A_1^a$, $A_2^a$, $\Pi_1^a$ and $\Pi_2^a$. The gauge field $A_0^a$ can be solved as
\begin{align}
\label{eqn:A0}
    (\partial_3)^2 A_0^a = \sum_{i=1,2} \sum_{b,c} (\partial_i \Pi_i^a  + g f^{abc}A_i^b \Pi_i^c) + gJ^{0a} \equiv Q^a \,,
\end{align}
where $i$ is only summed over $1$ and $2$ for the independent variables. To avoid confusion on the summation range, we will be explicit in summation signs from now on. Repeated indices will not be summed throughout the remainder of the paper.

Finally, the Hamiltonian density in terms of the independent variables can be written as
\begin{align}
    \mathcal{H} &= \sum_{a,i=1,2} \bigg( \frac{1}{2} \Pi_i^a \Pi_i^a + \frac{1}{2}(\partial_3A_i^a)(\partial_3A_i^a) -gA_i^a J^{ia} \bigg) + \frac{1}{2}\sum_{a} F_{12}^a F_{12}^a + \frac{1}{2} \sum_{a} (\partial_3A_0^a)(\partial_3A_0^a) \nn\\
    &- \sum_f\sum_{j=1,2,3} \bar{\psi}_f(i\gamma^j \partial_j - m_f) \psi_f \,,
\end{align}
where the $A_0^a$ field is determined by the independent variables as in Eq.~\eqref{eqn:A0}. The independent gauge field variables obey the standard commutation relations
\begin{subequations}
\label{eqn:commu}
\bea
[A_i^a(\boldsymbol x), A_j^b(\boldsymbol y)] &= 0 \,, \\
[\Pi_{i}^a(\boldsymbol x), \Pi_{j}^b(\boldsymbol y)] &= 0 \,,\\
\label{eqn:commu_APi}
[A_i^a(\boldsymbol x), \Pi_{j}^b(\boldsymbol y)] &= i \delta_{ij} \delta^{ab} \delta^{(3)}({\boldsymbol x} - {\boldsymbol y}) \,,
\eea
\end{subequations}
for $i,j\in\{1,2\}$. The fermion fields follow the standard anticommutation relations
\begin{subequations}
\label{eqn:anticommu}
\bea
\{\psi_{f\alpha}^i(\boldsymbol x), \psi_{f'\beta}^j(\boldsymbol y)\} &= 0\,, \\
\{\psi^{i \dagger}_{f\alpha}(\boldsymbol x), \psi^{j \dagger}_{f'\beta}(\boldsymbol y)\} &= 0\,, \\
\{\psi_{f \alpha}^i(\boldsymbol x), \psi^{j \dagger}_{f'\beta}(\boldsymbol y)\} &= \delta_{ff'}\delta_{\alpha\beta} \delta^{ij}\delta^{(3)}(\boldsymbol x-\boldsymbol y) \,,
\eea
\end{subequations}
where the subscripts $\alpha,\beta$ and the superscripts $i,j$ label the Dirac\footnote{There are four Dirac indices, labeling the fermion or antifermion field and its spin.} and the fundamental color indices, respectively. For example, expanding $\bar{\psi}_f \gamma^\mu T^a \psi_f$ for one flavor gives $\bar{\psi}^i_{f\alpha} \gamma_{\alpha\beta}^\mu (T^a)^{ij} \psi_{f \beta}^j$.

The $A_0^a$ interaction term can be written out more explicitly by using the Green's function $G(x_3, x_3')$ of $(\partial_3)^2$, which is defined by
\begin{align}
    (\partial_3)^2 G(x_3, x_3') = \delta(x_3 - x_3')\,,
\end{align}
and has the general solution
\begin{align}
    G(x_3, x_3') = \frac{1}{2}|x_3 - x_3'| + c \,,
\end{align}
where $c$ is a constant determined by the boundary condition. If we demand $\lim_{|x_3-x_3'|\to \infty} \allowbreak G(x_3, x_3') = 0$ then $c$ is infinite and the interaction strength diverges at $x_3=x_3'$, which is similar to the divergence of the Coulomb interaction at zero separation in QED. The divergence can be regulated on a lattice, which will be discussed in detail in Sec.~\ref{sec:dd}. With the Green's function and assuming all fields vanish at infinity, the $A_0^a$ interaction term in the Hamiltonian can be written as
\begin{align}
\label{eqn:QGQ}
    &\int {\rm d}^3 x \frac{1}{2} \sum_a [\partial_3A_0^a({\boldsymbol x})][\partial_3A_0^a({\boldsymbol x})] \nn\\
    &= -\frac{1}{2} \int {\rm d}x_1 \int {\rm d}x_2 \int {\rm d}x_3 \int {\rm d}x_3' \sum_a Q^a(x_1,x_2,x_3) G(x_3, x_3') Q^a(x_1,x_2,x_3') \,,
\end{align}
where the ``color charge'' $Q^a$ is defined in Eq.~\eqref{eqn:A0}.

\section{Discretization and Digitization in Field Basis}
\label{sec:dd}
We now discuss discretization of the Hamiltonian on a spatial lattice. 
We will express all quantities in units of the lattice spacing, which is equivalent to setting the lattice spacing to be unity. The continuum limit is taken by tuning the lattice coupling $g\to0$ and extrapolating physical observables. The physical value of the lattice spacing at a given $g$ is determined by scale fixing, which is a standard procedure in Euclidean lattice QCD computation and can be applied here too. We consider a cubic spatial lattice of size $L\in \mathbb{Z}_{>0}$ in each spatial direction with a volume $V=L^3$. Grid points on this lattice are specified by
\bea
\boldsymbol x = (x_1, x_2, x_3)\,,\ x_i\in \{1,2,3,\cdots,L \} \,.
\eea
We will impose the Dirichlet boundary condition throughout the remainder of this paper, in which all fields vanish outside the boundary of the cubic lattice. The finite lattice with the Dirichlet boundary condition removes the contribution from the zero momentum mode along the $x_3$ direction. Observables that extend to infinity may need special treatment in the boundary condition, as the zero mode contribution becomes crucial~\cite{Scheihing-Hitschfeld:2022xqx}; see Appendix~\ref{app}. Under the Dirichlet boundary condition with vanishing fields at $x_i=0$ or $x_i=L+1$, one can define discrete momenta on the lattice specified by
\bea
\label{eqn:discrete_momentum_Dirichlet}
\boldsymbol p = \frac{\pi}{(L+1)}(k_1,k_2,k_3) \,,\ k_i\in \{1,2,3,\cdots,L \} \,.
\eea

\subsection{Lattice QCD Hamiltonian in Axial Gauge}
To write the axial gauge lattice QCD Hamiltonian out explicitly, we follow Ref.~\cite{Yao:2025uxz} and introduce some notations for discrete differences on the lattice
\begin{subequations}
\bea
\partial_i^{(L+)} f({\boldsymbol x}) &\equiv  f({\boldsymbol x}+\hat{i}) - f({\boldsymbol x}) \,,\\
\partial_i^{(L-)} f({\boldsymbol x}) &\equiv  f({\boldsymbol x}) - f({\boldsymbol x}-\hat{i}) \,,
\eea
\end{subequations}
where $\hat{i}$ denotes a unit vector along the $i$ axis. The superscripts $(L\pm)$ differentiate how the finite difference is taken on the lattice. Under the Dirichlet boundary condition, these finite differences obey a lattice version of the ``integration by parts''~\cite{Yao:2025uxz}.

We now define the discrete version of the $(\partial_i)^2$ operator as
\bea
\label{eqn:discrete_partial_3^2}
[\partial_i^{(L)} ]^2 \equiv \partial_i^{(L+)} \partial_i^{(L-)} = \partial_i^{(L-)} \partial_i^{(L+)} \,,
\eea
where there is no summation over $i$. It acts on a function $f$ at site $\boldsymbol{x}=(x_1,x_2,x_3)$ as
\bea
[\partial_3^{(L)} ]^2 f({\boldsymbol{x}}) = f(x_1,x_2,x_3+1) + f(x_1,x_2,x_3-1) - 2 f(x_1,x_2,x_3) \,.
\eea
Using the method discussed in Ref.~\cite{chung2000discrete}, we find that the Green's function of the discrete operator $[\partial_3^{(L)}]^2$ is
\bea
\label{eqn:Green for discrete partial_3^2}
G(x_3, x_3') = \sum_{p_3} \frac{\phi_{p_3}(x_3)\phi_{p_3}(x_3') }{\lambda_{p_3}} \,,
\eea
where $\sum_{p_3}$ means $\sum_{k_3}$ with $k_3\in \{1,2,3,\cdots, L \}$ as in Eq.~\eqref{eqn:discrete_momentum_Dirichlet}. The eigenvalues and orthonormal eigenvectors of the operator $[\partial_3^{(L)}]^2$ are 
\begin{subequations}
\bea
\lambda_{p_3} &= 2\cos p_3 - 2 = 2\cos\frac{k_3\pi}{L+1} - 2 \,,\\
\phi_{p_3}(x_3) &= \sqrt{\frac{2}{L+1}}\sin(p_3 x_3) = \sqrt{\frac{2}{L+1}}\sin\frac{k_3 x_3 \pi}{L+1} \,. \label{eqn:phi_k(x)}
\eea
\end{subequations}
One can easily see that the Green's function is symmetric in its arguments $G(x_3,x_3') = G(x_3', x_3)$ and verify
\begin{align}
    [\partial_3^{(L)} ]^2 G(x_3, x_3') = \delta_{x_3x_3'} \,,
\end{align}
in which $\delta_{x_3 x_3'}$ is the Kronecker delta function. 

With these preparations, we are now ready to explicitly write out the lattice QCD Hamiltonian in axial gauge
\begin{subequations}
\label{eqn:latticeH}
\bea
H &= H_\Pi + H_A + H_{Q} + H_I + H_f \,, \\
\label{eqn:latticeHpi}
H_\Pi &= \frac{1}{2} \sum_{{\boldsymbol x},a} \sum_{i=1,2} [\Pi_i^a({\boldsymbol x})]^2 \,, \\
\label{eqn:latticeHA}
H_A &= \frac{1}{2}\sum_{{\boldsymbol x},a} \sum_{i=1,2} [\partial_3^{(L+)} A_i^a({\boldsymbol x}) ]^2 + \frac{1}{2} \sum_{{\boldsymbol x},a,b,c} [ \partial_1^{(L+)} A_2^a({\boldsymbol x}) - \partial_2^{(L+)} A_1^a({\boldsymbol x}) + gf^{abc}A_1^b({\boldsymbol x})A_2^c({\boldsymbol x}) ]^2 \,, \\
H_Q &= -\frac{1}{2} \sum_{{\boldsymbol x},a} \sum_{x_3'} Q^a({\boldsymbol x}) G(x_3,x_3') Q^a({\boldsymbol x}') \,,\\
\label{eqn:latticeHI}
H_I &= - g\sum_{{\boldsymbol x},a,f} \sum_{i=1,2} A_i^a({\boldsymbol x}) \bar{\psi}_f({\boldsymbol x}) \gamma^i T^a \psi_f({\boldsymbol x}) \,,\\
\label{eqn:latticeHf}
H_f &= - \frac{i}{2} \sum_{{\boldsymbol x},f} \sum_{j=1,2,3} \bar{\psi}_f({\boldsymbol x}) \gamma^j [ \psi_f({\boldsymbol x}+\hat{j}) - \psi_f({\boldsymbol x}-\hat{j}) ] + \sum_{{\boldsymbol x},f} m_f \bar{\psi}_f({\boldsymbol x}) \psi_f({\boldsymbol x}) \nn\\
&\quad -\frac{r}{2} \sum_{{\boldsymbol x},f} \sum_{j=1,2,3} \bar{\psi}_f({\boldsymbol x}) [\partial_j^{(L)} ]^2 \psi_f({\boldsymbol x}) \,,
\eea
\end{subequations}
where $\boldsymbol{x}=(x_1,x_2,x_3)$ and $\boldsymbol{x}'=(x_1,x_2,x_3')$, and we have added a Wilson term with the unitless positive coefficient $r>0$ to avoid the fermion doubling problem~\cite{Jordan:2014tma}. 
Unlike the Coulomb interaction term $H_C$ in the Coulomb gauge QED~\cite{Yao:2025uxz}, which is a double volume summation over six spatial coordinates, the $H_Q$ interaction term here only sums over four spatial coordinates.
Expanding $H_Q$ out using Eq.~\eqref{eqn:A0} gives
\begin{subequations}
\begin{align}
    H_Q &= H_{\Pi\Pi} + H_{A \Pi A \Pi} + H_{\Pi A \Pi} + H_{\Pi f} + H_{A \Pi f} + H_{ff} \,,\\
    \label{eqn:latticeHpipi}
    H_{\Pi\Pi} &= -\frac{1}{2} \sum_{{\boldsymbol x},a} \sum_{x_3'} \Big[\sum_{i=1,2} \partial_i^{(L+)} \Pi_i^a({\boldsymbol x}) \Big] G(x_3,x_3') \Big[\sum_{j=1,2} \partial_j^{(L+)} \Pi_j^a({\boldsymbol x}') \Big] \,,\\
    \label{eqn:latticeHApiApi}
    H_{A \Pi A \Pi} &= -\frac{g^2}{2} \sum_{{\boldsymbol x},a,b,c,d,e} \sum_{x_3'} \Big[\sum_{i=1,2} f^{abc} A_i^b({\boldsymbol x}) \Pi_i^c({\boldsymbol x}) \Big] G(x_3,x_3') \Big[\sum_{j=1,2} f^{ade} A_j^d({\boldsymbol x}')  \Pi_j^e({\boldsymbol x}') \Big] \,,\\
    \label{eqn:latticeHpiApi}
    H_{\Pi A \Pi} &= -g \sum_{{\boldsymbol x},a,b,c} \sum_{x_3'} \Big[\sum_{i=1,2} \partial_i^{(L+)} \Pi_i^a({\boldsymbol x}) \Big] G(x_3,x_3') \Big[\sum_{j=1,2} f^{abc} A_j^b({\boldsymbol x}')  \Pi_j^c({\boldsymbol x}') \Big] \,,\\
    \label{eqn:latticeHpif}
    H_{\Pi f} &= - g \sum_{{\boldsymbol x},a} \sum_{x_3'} \Big[\sum_{i=1,2} \partial_i^{(L+)} \Pi_i^a({\boldsymbol x}) \Big] G(x_3,x_3') \Big[\sum_f \bar{\psi}_f({\boldsymbol x}') \gamma^0 T^a \psi_f({\boldsymbol x}') \Big] \,,\\
    \label{eqn:latticeHApif}
    H_{A \Pi f} &= -g^2 \sum_{{\boldsymbol x},a,b,c} \sum_{x_3'} \Big[\sum_{i=1,2} f^{abc} A_i^b({\boldsymbol x}) \Pi_i^c({\boldsymbol x}) \Big] G(x_3,x_3') \Big[\sum_f \bar{\psi}_f({\boldsymbol x}') \gamma^0 T^a \psi_f({\boldsymbol x}') \Big] \,,\\
    \label{eqn:latticeHff}
    H_{ff} &= -\frac{g^2}{2} \sum_{{\boldsymbol x},a} \sum_{x_3'} \Big[\sum_f \bar{\psi}_f({\boldsymbol x}) \gamma^0 T^a \psi_f({\boldsymbol x}) \Big] G(x_3,x_3') \Big[\sum_{f'} \bar{\psi}_{f'}({\boldsymbol x}') \gamma^0 T^a \psi_{f'}({\boldsymbol x}') \Big] \,,
\end{align}
\end{subequations}
where again $\boldsymbol{x}=(x_1,x_2,x_3)$ and $\boldsymbol{x}'=(x_1,x_2,x_3')$, and we have used $G(x_3,x_3') = G(x_3',x_3)$. The independent field variables $A_1^a$, $A_2^a$, $\Pi_1^a$, $\Pi_2^a$, $\psi_{f\alpha}^i$, and $\psi_{f\alpha}^{i\dagger}$ on the lattice follow the same commutation or anticommutation relations as in Eqs.~\eqref{eqn:commu} and~\eqref{eqn:anticommu}, except for the Dirac delta functions $\delta^{(3)}(\boldsymbol{x} - \boldsymbol{y})$ for the continuum being replaced by the Kronecker delta functions $\delta_{\boldsymbol{x}\boldsymbol{y}}$ for the lattice.

We close this subsection by commenting that since the Hamiltonian $H$ does not contain any $\Pi_3^a$, the gauge condition $A_3^a=0$ is trivially maintained in time evolution, even under Trotterization. 

\subsection{Field Basis for Gauge Bosons and Map onto Qubits}
We represent the independent gauge fields in the local field basis $| A^a_i(\boldsymbol{x})\rangle$, defined as
\bea
\hat{A}^a_i({\boldsymbol x}) |A^b_j({\boldsymbol y})\rangle = \delta_{ij} \delta^{ab} \delta_{{\boldsymbol x}{\boldsymbol y}} A^b_j({\boldsymbol y}) |A^b_j({\boldsymbol y})\rangle \,,
\eea
for $i,j\in\{1,2\}$, where to distinguish the operator from the field value, we added a hat to the operator. In the following, we will not make this distinction. 
The total Hilbert space over the whole lattice is a tensor product
\bea
\bigotimes_{{\boldsymbol x},a,i=1,2} |A^a_i({\boldsymbol x})\rangle \,.
\eea
In practical numerical calculations, one must digitize and truncate the gauge field values, which we take to be in the interval $[-A_{\rm max}, A_{\rm max}]$ with the spacing $\delta \!A$ for all $\boldsymbol{x}, a, i$. All gauge field basis states can be mapped onto qubits just as in the scalar field case. The total number of qubits needed is $16 n_A V$ for eight adjoint colors and two spatial indices. The differences from the case of lattice QED in Coulomb gauge are the eight adjoint colors and that there are only two spatial indices rather than three. The number of qubits needed at one site for one color and one spatial index is
\bea
n_A = \lceil \log_2(2 A_{\rm max}/\delta\!A+1) \rceil \,.
\eea
In Sec.~\ref{sec:qubit_cost}, we will use the wavefunction fidelity amplitude up to a given energy and prove bounds on $A_{\rm max}$ and $\delta\!A$, which together give a bound on $n_A$.

\subsection{Field Basis for Fermions and Map onto Qubits}
Similar to the cases of pure fermion theory~\cite{Jordan:2014tma} and lattice QED in Coulomb gauge~\cite{Yao:2025uxz}, we specify the fermion field basis by using the following commuting observables
\bea
\label{eqn:S_x}
S = \{ \psi^{i\dagger}_{f\alpha}({\boldsymbol x}) \psi_{f\alpha}^i({\boldsymbol x}) \ |\ {\boldsymbol x},\, f,\, i=1,2,3,\, \alpha=1,2,3,4 \} \,,
\eea
where repeated indices are not summed. As before, $f$ labels the flavor, $i$ is the fundamental color index, and $\alpha$ denotes the Dirac index. Because of the anticommutation relations in Eq.~\eqref{eqn:anticommu}, all the operators in $S$ commute with each other and each has eigenvalues $0$ and $1$. The local fermion Hilbert space for a given $f,i,\alpha$ at one lattice site has dimension $2$ and the basis is specified by the eigenvalues of $\psi^{i\dagger}_{f\alpha}({\boldsymbol x}) \psi^i_{f\alpha}({\boldsymbol x})$. It can be mapped onto a qubit naturally with the eigenbasis corresponding to the eigenvalue $n_{f\alpha}^i$ mapped onto the qubit state $|n_{f\alpha}^i\rangle$ for $n_{f\alpha}^i=0,1$. The total number of commuting observables in $S$ is $12n_f V$ for $n_f$ flavors, three fundamental colors, and four Dirac indices, requiring $12n_f V$ qubits to represent all the fermionic degrees of freedom.

We will discuss in Sec.~\ref{sec:time_and_gate} how to maintain the anticommutation relations between the fermion fields when implementing the Hamiltonian on a quantum computer.

\section{Bound on Qubit Cost}
\label{sec:qubit_cost}
In this section we prove a bound on the total number of qubits needed for describing all physical states up to an energy $E$ on a given lattice setup with an accuracy $\epsilon$. The reason why we only care about states up to a given energy is that on any finite lattice with any nonzero lattice spacing, states close to the middle of the spectrum and above are just lattice artifacts. The proof uses some tricks developed in Ref.~\cite{Jordan:2011ci} for lattice scalar fields and in Ref.~\cite{Yao:2025uxz} for lattice QED in Coulomb gauge. We also derive some new formulas for the axial gauge setup. The results are summarized in Sec.~\ref{sec:total num qubits}.

The complete field basis can be written as
\bea
\Big\{ \bigotimes_{{\boldsymbol x},a,i,f,j,\alpha} \big[ |A^a_i({\boldsymbol x}) \rangle  \otimes  |n_{f\alpha}^j({\boldsymbol x})\rangle \big] \Big\} \,,
\eea
spanning all possible field values, 
where $i\in\{1,2\}$ and $j\in\{1,2,3\}$. An arbitrary state $|\Psi \rangle$ with energy $E$ can be represented in this complete basis as
\bea
|\Psi \rangle &= \bigotimes_{{\boldsymbol x},a,i,f,j,\alpha} \bigg[ \int_{-\infty}^{\infty} {\rm d}A^a_i({\boldsymbol x}) |A^a_i({\boldsymbol x}) \rangle  \otimes \!\! \sum_{n^j_{f\alpha}\!({\boldsymbol x})=0}^1 \!\! |n_{f\alpha}^j({\boldsymbol x})\rangle \bigg] \Psi[ \{A^a_i({\boldsymbol x})\}; \{n_{f\alpha}^j({\boldsymbol x})\} ] \,,
\eea
in which $\Psi[ \{A^a_i({\boldsymbol x})\}; \{n_{f\alpha}^j({\boldsymbol x})\} ]$ is the probability amplitude of being in the corresponding basis state. If the gauge field value is truncated to the interval $[-A_{\rm max}, A_{\rm max}]$, the truncated wavefunction becomes
\bea
|\Psi_{\rm cut} \rangle &= \bigotimes_{{\boldsymbol x},a,i,f,j,\alpha} \bigg[ \int_{-A_{\rm max}}^{A_{\rm max}} {\rm d}A^a_i({\boldsymbol x}) |A^a_i({\boldsymbol x}) \rangle  \otimes \!\! \sum_{n^j_{f\alpha}\!({\boldsymbol x})=0}^1 \!\! |n_{f\alpha}^j({\boldsymbol x})\rangle \bigg] \Psi[ \{A^a_i({\boldsymbol x})\}; \{n_{f\alpha}^j({\boldsymbol x})\} ] \,.
\eea
Their wavefunction overlap (fidelity amplitude) is
\bea
\langle \Psi | \Psi_{\rm cut} \rangle = \prod_{{\boldsymbol x},a,i,f,j,\alpha} \bigg[ \int_{-A_{\rm max}}^{A_{\rm max}} {\rm d}A_i({\boldsymbol x}) \sum_{n^j_{f\alpha}({\boldsymbol x})=0}^1 \bigg] \big| \Psi[ \{{A}_i^a({\boldsymbol x})\}; \{n_{f\alpha}^j({\boldsymbol x})\} ] \big|^2 \,.
\eea
We use $P_{\rm out}({\bs x},a, i) \equiv P(|A^a_i({\bs x})| > A_{\rm max})$ to denote the probability of the state $|\Psi\rangle$ having a field value $A_i^a$ whose magnitude is greater than $A_{\rm max}$ at ${\bs x}$.
Then through the probability of a union of events~\cite{Jordan:2011ci}, we find
\bea
\label{eqn:psi_psicut_bound}
\langle \Psi | \Psi_{\rm cut} \rangle \geq 1- \sum_{{\bs x}, a, i} P_{\rm out}({\bs x}, a, i) \geq 1- 16V \max_{{\bs x}, a, i}P_{\rm out}({\bs x}, a, i) \,,
\eea
where the factor of $16$ comes from eight adjoint colors and two spatial indices.
If we want an accuracy $\epsilon$, i.e., $\langle \Psi | \Psi_{\rm cut} \rangle > 1-\epsilon$, we will require
\bea
\label{eqn:Pmax_epsilon}
\max_{{\bs x}, a, i}P_{\rm out}({\bs x}, a, i) \leq \frac{\epsilon}{16{V}} \,.
\eea

The remaining proof will use the Chebyshev inequality to express $\max_{{\bs x}, a,i}P_{\rm out}({\bs x}, a, i)$ in terms of the expected value of some field operator in the Hamiltonian, which can be bounded by the state energy $E$. To bound the expected value of the relevant field operator, we first need to shift the Hamiltonian properly and decompose it into positive semidefinite parts, as done for the lattice QED in Coulomb gauge~\cite{Yao:2025uxz}.

\subsection{Positive Semidefinite Hamiltonian}
From the lattice Hamiltonian in Eq.~\eqref{eqn:latticeH}, we note that $H_\Pi$, $H_{A}$ and $H_Q$ are already positive semidefinite: $H_\Pi$ and $H_{A}$ are written in terms of squares with positive coefficients and $H_Q$ can be written as by using the Green's function 
\bea
H_Q = \sum_{x_1,x_2,p_3,a} \frac{1}{8\sin^2(p_3/2)} \Big[\sum_{x_3} \phi_{p_3}(x_3) Q^a(x_1,x_2,x_3) \Big]^2 \,.
\eea
So we only need to deal with the fermion interaction and free parts, $H_I$ and $H_f$. To this end, we introduce a discrete Fourier-like transform for operators and functions
\begin{subequations}
\bea
\mathcal{O}(x_1,x_2,x_3) &= \sum_{p_3} \phi_{p_3}(x_3) \mathcal{O}(x_1,x_2,p_3) \,,\\
\mathcal{O}(x_1,x_2,p_3) &= \sum_{x_3} \phi_{p_3}(x_3) \mathcal{O}(x_1,x_2,x_3) \,,
\eea
\end{subequations}
where $\phi_{p}(x)$ is given in Eq.~\eqref{eqn:phi_k(x)}. If $\mathcal{O}({\boldsymbol x}) = \mathcal{O}(x_1,x_2,x_3)$ is Hermitian, then $\mathcal{O}(x_1,x_2,p_3)$ is also Hermitian since $\phi_{p_3}(x_3)$ is a real function. 

Adding the first term of $H_A$ as in Eq.~\eqref{eqn:latticeHA} and $H_I$ in Eq.~\eqref{eqn:latticeHI} gives
\begin{align}
\label{eqn:complete_square}
  &  \frac{1}{2}\sum_{{\boldsymbol x},a} \sum_{i=1,2} [\partial_3^{(L+)} A_i^a({\boldsymbol x}) ]^2 - g\sum_{{\boldsymbol x},a,f} \sum_{i=1,2} A_i^a({\boldsymbol x}) \bar{\psi}_f({\boldsymbol x}) \gamma^i T^a \psi_f({\boldsymbol x})  \nn\\
 =\ & - \frac{1}{2}\sum_{{\boldsymbol x},a} \sum_{i=1,2} \{ A_i^a({\boldsymbol x}) [ \partial_3^{(L)} ]^2 A_i^a({\boldsymbol x}) +2 g A_i^a({\boldsymbol x}) J^{ia}({\boldsymbol x}) \} \nn\\
 =\ & \!\! \sum_{x_1,x_2,p_3,a}\sum_{i=1,2} \bigg\{ 2\sin^2\!\left(\frac{p_3}{2}\right) \!\left[ A_i^a(x_1,x_2,p_3) - \frac{gJ^{ia}(x_1,x_2,p_3)}{4\sin^2(p_3/2)} \right]^2 - \frac{g^2 [J^{ia}(x_1,x_2,p_3)]^2 }{8\sin^2(p_3/2)} \bigg\} \,,
\end{align}
in which the first term after the last equal sign is positive semidefinite and the second term can be bounded as
\bea
\label{eqn:bound_dE1_analysis}
\sum_{x_1,x_2,p_3,a}\sum_{i=1,2} \frac{g^2 [J^{ia}(x_1,x_2,p_3)]^2 }{8\sin^2(p_3/2)} & \leq \frac{g^2 }{8\sin^2[\pi/(2L+2)]} \sum_{\boldsymbol{x},a}\sum_{i=1,2} [J^{ia}(\boldsymbol{x})]^2 \nn\\
& \leq \frac{2g^2 V}{\sin^2[\pi/(2L+2)]} \big[ \max_{\boldsymbol{x},a,i} ||J^{ia}(\boldsymbol{x})|| \big]^2 \,,
\eea
for eight adjoint colors and two spatial indices. The operator norm $||J^{ia}(\boldsymbol{x})||$ can be bounded similarly as done in the Coulomb gauge QED case~\cite{Yao:2025uxz}. For one flavor at one site, we have
\bea
|| \bar{\psi}_f \gamma^i T^a \psi_f || = 2 ||T^a||_1 \,, 
\eea
where $||T^a||_1$ is the trace norm defined by the sum of all the eigenvalue magnitudes of the matrix. For $a=1,2,\cdots,7$, $||T^a||_1=1$ while for $a=8$, $||T^a||_1=2/\sqrt{3}$. Altogether we find
\bea
\sum_{x_1,x_2,p_3,a}\sum_{i=1,2} \frac{g^2 [J^{ia}(x_1,x_2,p_3)]^2 }{8\sin^2(p_3/2)} \leq \frac{32g^2 n_f^2 V}{3\sin^2[\pi/(2L+2)]} \approx \frac{128g^2 n_f^2 V^{5/3}}{3\pi^2} \,,
\eea
where the approximation sign is obtained by assuming that $L$ is large. A tighter bound can be obtained by analyzing $\max_{\boldsymbol{x}} ||\sum_{a,i=1,2}J^{ia}(\boldsymbol{x})J^{ia}(\boldsymbol{x})||$ directly in Eq.~\eqref{eqn:bound_dE1_analysis}. We expect this to improve the bound by a constant factor but not to change the scalings with $V$ and $g$.

We can then add to the Hamiltonian a constant 
\bea
\Delta E_1 = \frac{32g^2 n_f^2 V}{3\sin^2[\pi/(2L+2)]} \,,
\eea
which when combined with the last term in Eq.~\eqref{eqn:complete_square} gives a positive semidefinite term. 

The treatment of the free fermion part $H_f$ is similar to that in the Coulomb gauge QED case~\cite{Yao:2025uxz}. $H_f$ is not positive semidefinite because of the Dirac sea of negative-energy particles. It can be made positive semidefinite by adding a constant energy proportional to the volume
\bea
\Delta E_2 = 6V \sum_f \sqrt{3+ m_f^2 + 12m_f r + 36 r^2} \,,
\eea
where the factor six originates from two spins and three fundamental colors. This shift was first derived in Ref.~\cite{Yao:2025uxz} and later used in a different Coulomb gauge lattice QED study~\cite{Li:2024ide} (see its Supplemental Material).

The Hamiltonian $H+\Delta E_1 + \Delta E_2$ can be decomposed into positive semidefinite pieces, which can then be used to bound expected values of certain operators in the Hamiltonian in terms of the state energy and accuracy. We now prove these bounds that eventually lead to a bound on the number of qubits. 

\subsection{Bound on $A_{\rm max}$}
We use the Chebyshev inequality, which states that any random variable $X$ with finite mean $\mu_X$ and variance $\sigma_X^2$ has a bounded probability of deviating from its mean
\bea
\label{eqn:X_chebyshev}
P(|X- \mu_X| > \kappa \sigma_X) < \frac{1}{\kappa^2} \,,
\eea
where $\kappa>0$. In the following, we will use Proposition 2 of Ref.~\cite{Jordan:2011ci}, which states
\begin{subequations}
\label{eqn:mu_sigma_X}
\bea
\mu_X &= \langle \Psi | X | \Psi \rangle \leq \sqrt{\langle \Psi | X^2 | \Psi \rangle} \,,\\
\sigma_X &= \sqrt{ \langle \Psi | [X - \mu_X]^2 | \Psi \rangle } \leq \sqrt{\langle \Psi | X^2 | \Psi \rangle} \,,
\eea
\end{subequations}
and leads to
\bea
P\big(|X| > (\kappa+1) \sqrt{\langle \Psi | X^2 | \Psi \rangle} \big) < \frac{1}{\kappa^2} \,.
\eea

We consider a state $|\Psi\rangle$ with energy $E$. Using the second-to-last term in Eq.~\eqref{eqn:complete_square}, which is positive semidefinite, we obtain
\bea
\label{eqn:shift_E}
E'\equiv E+\Delta E_1+\Delta E_2 & \geq 
\langle \Psi | \sum_{x_1,x_2,p_3,a}\sum_{i=1,2} 2\sin^2\!\left(\frac{p_3}{2}\right) \!\left[ A_i^a(x_1,x_2,p_3) - \frac{gJ^{ia}(x_1,x_2,p_3)}{4\sin^2(p_3/2)} \right]^2   | \Psi \rangle \nn\\
& \geq 2\sin^2\!\left(\frac{\pi}{2L+2}\right) \langle \Psi | \{ A_i^a(\boldsymbol{x}) + g [\partial_3^{(L)}]^{-2} J^{ia}(\boldsymbol{x}) \}^2 |  \Psi \rangle \,,
\eea
which is true $\forall \boldsymbol{x}, a, i$. For a translationally invariant state, one can tighten the bound by adding a multiplicative volume factor $V$ in the last term, as demonstrated in the scalar field theory case~\cite{Yang:2026zpa}. However, many interesting real-time lattice calculations involve initial states that break translational invariance significantly. For example, in the calculation of the energy-energy correlator, a perturbation operator is applied at one spatial site on top of the vacuum so almost all excitation energy sits at one spatial point~\cite{Lee:2024jnt}. The more conservative bound derived here without the volume factor still applies to this extreme case. In any case, one can easily add the volume factor when applying the bound to translationally invariant setups.

Using the Chebyshev inequality and setting
\bea
\kappa = \sqrt{\frac{16V}{\epsilon}}\,, \quad X_{\rm max} = (\kappa+1) \sqrt{\frac{E'}{2\sin^2[\pi/(2L+2)]}}\,,
\eea
we obtain
\bea
P(| A_i^a(\boldsymbol{x}) + g [\partial_3^{(L)}]^{-2} J^{ia}(\boldsymbol{x}) |>X_{\rm max}) < \frac{\epsilon}{16V} \,, \forall \boldsymbol{x}, a, i\,.
\eea
To obtain a final bound on $A_{\rm max}$, we use
\bea
| A_i^a(\boldsymbol{x})| &\leq | A_i^a(\boldsymbol{x}) + g [\partial_3^{(L)}]^{-2} J^{ia}(\boldsymbol{x}) |  + |g [\partial_3^{(L)}]^{-2} J^{ia}(\boldsymbol{x}) | \nn\\
&\leq | A_i^a(\boldsymbol{x}) + g [\partial_3^{(L)}]^{-2} J^{ia}(\boldsymbol{x}) | + \frac{g n_f}{\sqrt{3}\sin^2[\pi/(2L+2)]} \,.
\eea
We see that by choosing
\bea
\label{eqn:bound_A}
A_{\rm max} = X_{\rm max} + \frac{gn_f}{\sqrt{3}\sin^2[\pi/(2L+2)]} \approx \sqrt{\frac{32E'V^{5/3}}{\pi^2\epsilon}} + \frac{4g  n_fV^{2/3}}{\sqrt{3} \pi^2} \,,
\eea
we can achieve the bound on the wavefunction accuracy in Eq.~\eqref{eqn:Pmax_epsilon}, i.e.,
\bea
P(| A_i^a(\boldsymbol{x}) | > A_{\rm max}) < \frac{\epsilon}{16V} \,.
\eea

\subsection{Bound on $\delta\! A$}
Because of the commutation relation in Eq.~\eqref{eqn:commu_APi}, the conjugate variable space that is defined by\footnote{Here we added a hat to distinguish the operator from the conjugate field value, which we will not do in the following.}
\bea
\hat{\Pi}_i^a(\bs x)|{\Pi}_j^b(\bs y) \rangle = \delta_{ij}\delta^{ab}\delta_{{\bs x}{\bs y}}{\Pi}_j^b(\bs y)|{\Pi}_j^b(\bs y) \rangle \,,
\eea
for $i,j\in\{1,2\}$, can be transformed into from the field basis by a local Fourier transform~\cite{Jordan:2011ci,Jordan:2012xnu}. This gives
\bea
\label{eqn:Pimax_deltaA}
{\Pi}_{\rm max} = \frac{\pi}{\delta\!{A}} \,.
\eea
We can then use the electric energy $\Pi_i^a\Pi_i^a$ to bound $\delta\!{A}$, which starts with
\bea
E' \geq \langle \Psi | \frac{1}{2}{\Pi}^a_i({\boldsymbol x}) {\Pi}^a_i({\boldsymbol x})  | \Psi \rangle \,, \forall\boldsymbol{x},a,i\,. 
\eea
A similar usage of the Chebyshev inequality leads us to conclude that by choosing
\bea
\label{eqn:Pimax}
\Pi_{\rm max} = \sqrt{\frac{32E'V}{\epsilon}} \,,
\eea
we can guarantee
\bea
P(|\Pi^a_i({\boldsymbol x})|>\Pi_{\rm max}) < \frac{\epsilon}{16V} \,, \forall\boldsymbol{x},a,i\,,
\eea
which again guarantees the accuracy in Eq.~\eqref{eqn:Pmax_epsilon}.

\subsection{Total Number of Qubits}
\label{sec:total num qubits}
To describe all states up to a given energy $E$ on a lattice with size $V$, coupling $g$, and $n_f$ flavors, the total number of qubits needed is
\bea
16n_AV + 12n_f V \,,
\eea
where $n_A$ is bounded from Eqs.~\eqref{eqn:bound_A}, \eqref{eqn:Pimax_deltaA}, and~\eqref{eqn:Pimax} as
\bea
\label{eqn:nA_final}
n_A \approx \log_2\bigg( \frac{64 E' V^{4/3}}{\pi^2\epsilon} + \frac{32\sqrt{2}g n_f E'^{1/2} V^{7/6}}{\sqrt{3}\pi^3\epsilon^{1/2}} \bigg) \,,
\eea
where $E'$ is the shifted energy in Eq.~\eqref{eqn:shift_E}.

As a comparison, we also list the number of qubits needed for pure gauge theory, since non-Abelian gauge fields self interact and pure gauge theory itself is an interesting theory
\bea
n_A^{\rm pure\ gauge} \approx \log_2\bigg( \frac{64 E V^{4/3}}{\pi^2\epsilon} \bigg) \,,
\eea
where the energy is the unshifted one $E$.

\section{Quantum Algorithm for Time Evolution and Gate Cost}
\label{sec:time_and_gate}
The quantum algorithm for simulating the real-time evolution of the axial gauge lattice QCD shares some similarity with that for the Coulomb gauge lattice QED~\cite{Yao:2025uxz}, i.e., the Hamiltonian is Trotterized and the gauge degrees of freedom are represented in the truncated and digitized local field basis, which can be efficiently transformed into the local canonical conjugate variable basis by quantum Fourier transform. Here new technical developments are also needed due to the non-Abelian nature of the theory, which we will explain in detail in the next two subsections. We will focus on estimating the gate costs for CNOT and single-qubit rotation gates and neglect the costs for Hadamard and phase gates, as the CNOT gate count (describing the amount of entanglement needed) is the limiting quantum resource on near-term quantum devices while the number of single-qubit rotation gates is approximately proportional to the number of T gates by a factor of $0.02$--$0.1$~\cite{Halimeh:2024bth,Kliuchnikov:2014pzk,Campbell:2020wqh}, which describes the amount of nonstabilizerness needed and is the limiting quantum resource on fault-tolerant quantum computers. Gate costs for Trotterized Hamiltonian evolution are summarized in Sec.~\ref{sec:gate cost}.

\subsection{Pure Gauge Field Terms}
The gauge field value at each site is digitized into $2^{n_A}$ levels. These levels can be bijectively mapped to states of $n_A$ qubits. For example, $|-A_{\rm max}\rangle \to |000\cdots 00\rangle$, $|-A_{\rm max}+\delta \! A\rangle \to |000\cdots 01\rangle$, $|- A_{\rm max}+2\delta\! A\rangle \to |000\cdots 10\rangle$, $\cdots$, $|A_{\rm max} \rangle \to |111\cdots 11\rangle$. Then the local field operator $A_i^a(\boldsymbol{x})$ in this local basis can be decomposed in terms of Pauli operators as~\cite{Klco:2018zqz,Halimeh:2024bth}
\bea
\label{eqn:Adecomp}
A_i^a({\bs x}) = -\frac{1}{2}\delta\! A \sum_{n=0}^{n_A-1} 2^n \sigma^z_n({\bs x},a,i) \,,
\eea
where $n$ denotes the $n$th qubit representing the local gauge field. The argument of $\sigma^z_n({\bs x},a,i)$ emphasizes that the Pauli $Z$ matrix acts on the qubits for the gauge field at position ${\bs x}$ with color index $a$ and spatial index $i$. 

\subsubsection{$H_A$}
With Eq.~\eqref{eqn:Adecomp}, the time evolution driven by $H_A$ in Eq.~\eqref{eqn:latticeHA} can be written out analytically. Under the discrete ``integration by parts'', the first term in Eq.~\eqref{eqn:latticeHA} is
\bea
\label{eqn:A partial3_sqd A}
\frac{1}{2}\sum_{\boldsymbol{x},a,i} A_i^a(\boldsymbol{x}) [A_i^a(\boldsymbol{x}+\hat{x}_3) + A_i^a(\boldsymbol{x}-\hat{x}_3) -2A_i^a(\boldsymbol{x}) ] \,,
\eea
where $\hat{x}_3$ is a unit vector along the third spatial direction. To illustrate, we pick one term $\frac{1}{2}A_i^a(\boldsymbol{x}) A_i^a(\boldsymbol{x}+\hat{x}_3)$ (no summation over $i$ and $a$), and decompose its time evolution for a Trotter step $\Delta t$ as
\begin{align}
 e^{-i A_i^a(\boldsymbol{x}) A_i^a(\boldsymbol{x}+\hat{x}_3) \Delta t/2} &= \exp \bigg[ \frac{-i}{8}\delta\!A^2\Delta t \sum_{n_1=0}^{n_A-1} \sum_{n_2=0}^{n_A-1} 2^{n_1} 2^{n_2} \sigma^z_{n_1}({\bs x},a,i) \sigma^z_{n_2}({\bs x}+\hat{x}_3,a,i) \bigg] \nn\\
 &= \prod_{n_1=0}^{n_A-1} \prod_{n_2=0}^{n_A-1} \exp [ -i 2^{n_1+n_2-3} \delta\!A^2\Delta t \, \sigma^z_{n_1}({\bs x},a,i) \sigma^z_{n_2}({\bs x}+\hat{x}_3,a,i) ] \,.
\end{align}
Each term in the product can be decomposed into two CNOT gates and one single-qubit rotation gate. At one site, we have $n_A^2$ such terms for one color and one spatial index. In total, for the term in Eq.~\eqref{eqn:A partial3_sqd A}, we have $48 n_A^2 V$ such terms (eight colors, two spatial indices, and three quadratic pieces), and thus we need $96n_A^2 V$ CNOT gates and $48 n_A^2 V$ single-qubit rotation gates per Trotter step.

Next we study the second term of $H_A$ as in Eq.~\eqref{eqn:latticeHA}. In addition to quadratic terms, we also have cubic and quartic terms such as 
\begin{subequations}
\bea
&f^{abc} A_i^a(\boldsymbol{x}') A_1^b(\boldsymbol{x}) A_2^c(\boldsymbol{x})\,, \\
&f^{abc}f^{ade}A_1^b(\boldsymbol{x}) A_2^c(\boldsymbol{x}) \allowbreak A_1^d(\boldsymbol{x}) A_2^e(\boldsymbol{x}) \,,
\eea
\end{subequations}
respectively.
In the second term of $H_A$, we have $80V$ quadratic terms (24 from $A_1^a[\partial_2^{(L)}]^2A_1^a$, 24 from $A_2^a[\partial_1^{(L)}]^2A_2^a$ and $32$ from $\partial_1^{(L+)} A_2^a \partial_2^{(L+)}A_1^a$ at one site), $108 V$ cubic terms [54 nonzero $f^{abc}$ and $f^{abc}A_i^a(\boldsymbol{x}')A_1^b(\boldsymbol{x})A_2^c(\boldsymbol{x})$ vanishes when $\boldsymbol{x}'=\boldsymbol{x}$ and $i=1,2$], and $364V$ quartic terms (364 nonzero $\sum_a f^{abc}f^{ade}$). One quadratic operator is decomposed into a sum of $n_A^2$ tensor products of two Pauli $Z$ matrices, one cubic operator into $n_A^3$ products of three Pauli $Z$ matrices, and one quartic operator into $n_A^4$ products of four Pauli $Z$ matrices. The time evolution of a length-$l$ Pauli string takes $2l-2$ CNOT and a single-qubit rotation to implement per Trotter step, plus some Hadamard and phase gates that we will not count. In total, we need $160 n_A^2 V + 432 n_A^3V + 2184 n_A^4V$ CNOT gates and $ 80 n_A^2 V + 108 n_A^3V + 364 n_A^4V$ single-qubit rotation gates to implement one Trotter step for the second term of $H_A$.

\subsubsection{$H_\Pi$}
For the $H_\Pi$ term in Eq.~\eqref{eqn:latticeHpi}, we transform to the local canonical conjugate variable basis in which $H_\Pi$ is diagonal. The transformation is achieved by local quantum Fourier transform and can be performed in parallel for all sites, colors and spatial indices. At one site for one color and one spatial index, the quantum Fourier transform costs $n_A^2 - n_A + 3\lfloor n_A/2 \rfloor$ CNOT gates and $(n_A^2+n_A)/2$ single-qubit rotation gates. For the total cost of the transformation across the whole lattice, we then multiply by a factor of $16V$ (eight colors and two spatial indices). In the local canonical conjugate variable basis, the operator $\Pi_i^a(\boldsymbol{x})$ has a similar Pauli $Z$ decomposition
\bea
\label{eqn:pidecomp}
\Pi_i^a({\bs x}) = -\frac{1}{2}\delta \Pi \sum_{n'=0}^{n_A-1} 2^{n'} \sigma^z_{n'}({\bs x},a,i) \,,
\eea
where $\delta \Pi = \pi/A_{\rm max}$. 
The time evolution driven by $H_\Pi$ can be decomposed into a product of evolution operators consisting of two Pauli Z matrices. For one-step time evolution, we need $32n_A^2V$ CNOT gates and $16n_A^2 V$ single-qubit rotation gates.

\subsubsection{$H_{\Pi\Pi}$}
We then discuss the $H_{\Pi\Pi}$ term in Eq.~\eqref{eqn:latticeHpipi}. It can be implemented as the $H_\Pi$ term in the same local canonical conjugate variable basis, in which it is also diagonal. It contains $80V^{4/3}$ quadratic terms (24 from $\Pi_1^a[\partial_2^{(L)}]^2\Pi_1^a$, 24 from $\Pi_2^a[\partial_1^{(L)}]^2\Pi_2^a$ and $32$ from $\partial_1^{(L+)} \Pi_2^a \partial_2^{(L+)}\Pi_1^a$ at fixed positions), so we need $160n_A^2 V^{4/3}$ CNOT gates and $80n_A^2 V^{4/3}$ single-qubit rotation gates to implement one Trotter step driven by $H_{\Pi\Pi}$.

\subsubsection{$H_{ A\Pi A \Pi}$}
The $H_{A\Pi A\Pi }$ term in Eq.~\eqref{eqn:latticeHApiApi} involves a new technical subtlety that is absent in the Coulomb gauge lattice QED case. When $x_3\neq x_3'$, we see that none of the terms
\bea
\label{eqn:ffApiApi}
f^{abc} f^{ade} A_i^b(\boldsymbol{x}) \Pi_i^c(\boldsymbol{x}) A_j^d(\boldsymbol{x}') \Pi_j^e(\boldsymbol{x}')\,, 
\eea
contains both $A_i^a(\boldsymbol{x})$ and $\Pi_i^a(\boldsymbol{x})$ simultaneously at the same site with the same color and spatial indices. In other words, if we start in the local field basis, we can quantum Fourier transform for a subset of positions, color and spatial indices such that Eq.~\eqref{eqn:ffApiApi} is a sum of tensor products of four Pauli $Z$ matrices, i.e., each operator in Eq.~\eqref{eqn:ffApiApi} can be written as Eq.~\eqref{eqn:Adecomp} or~\eqref{eqn:pidecomp}, in which no Pauli $Z$ matrices act on the same qubit. After the proper quantum Fourier transform, $H_{A\Pi A\Pi}$ with $x_3\neq x_3'$ contains $1456 n_A^4 V(L-1)$ tensor products of four Pauli $Z$ matrices across the whole lattice. To implement one Trotter step, it takes $8736 n_A^4 V(L-1)$ CNOT and $1456 n_A^4 V(L-1)$ single-qubit rotation gates. 

The really tricky part is when $x_3 = x_3'$. Among $\sum_af^{abc}f^{ade}$ there are eight nonzero terms with $b=e$ but $c\neq d$, eight nonzero terms with $c=d$ but $b\neq e$, and 50 nonzero terms with $b=e$ and $c=d$. These lead to operators of the form
\bea
A_i^a(\boldsymbol{x}) \Pi_i^a(\boldsymbol{x}) A_i^b(\boldsymbol{x}) \Pi_i^c(\boldsymbol{x}) \,,
\eea
where $i$ and $a$ are not summed and both $b$ and $c$ are not equal to $a$ ($b$ and $c$ may be the same or not). The operator $A_i^a(\boldsymbol{x}) \Pi_i^a(\boldsymbol{x})$ is not diagonal in either the field basis or the canonical conjugate variable basis and thus does not have a straightforward decomposition into tensor products of Pauli Z matrices. Furthermore, the operator $A_i^a(\boldsymbol{x}) \Pi_i^a(\boldsymbol{x})$ itself has an ambiguity in the ordering due to the nonvanishing commutator in Eq.~\eqref{eqn:commu_APi}. However, as the original term ($H_{A\Pi A\Pi}$ with $x_3= x_3'$) is symmetric in the fields' arguments and indices, we expect $A_i^a(\boldsymbol{x}) \Pi_i^a(\boldsymbol{x})$ to always appear together with $\Pi_i^a(\boldsymbol{x}) A_i^a(\boldsymbol{x}) $ so there is no ambiguity in the ordering.

Finding the corresponding quantum circuit for $A_i^a(\boldsymbol{x}) \Pi_i^a(\boldsymbol{x}) + \Pi_i^a(\boldsymbol{x}) A_i^a(\boldsymbol{x}) $ (no summation over indices) is just like a quantum mechanical problem of implementing $xp+px$ in time evolution. Some algebra gives\footnote{These coefficients $a,b,c,d$ were found by Claude. The decomposition holds for any real $b\neq 0$ with $a=(1-e^{-2\Delta})/4b$, $c=(1-e^{2\Delta})/4b$, $d=-be^{-2\Delta}$; we take $b=\sqrt{|\Delta|/2}$ so that all coefficients are $O(\sqrt{|\Delta|})$.}
\bea
&e^{-i\Delta(xp+px)} = e^{-ia x^2} e^{-ib p^2} e^{-ic x^2} e^{-id p^2}\,, \nn\\
&a = \frac{1-e^{-2\Delta}}{2\sqrt{2|\Delta|}} \,,\quad b = \sqrt{\frac{|\Delta|}{2}} \,,\quad c = \frac{1-e^{2\Delta}}{2\sqrt{2|\Delta|}} \,,\quad d = -e^{-2\Delta}\sqrt{\frac{|\Delta|}{2}}\,.
\eea
This relation is exact. The time evolution driven by $[A_i^a(\boldsymbol{x}) \Pi_i^a(\boldsymbol{x}) + \Pi_i^a(\boldsymbol{x}) A_i^a(\boldsymbol{x})] A_i^b(\boldsymbol{x}) \Pi_i^c(\boldsymbol{x})$ can then be implemented as
\bea
& \exp\{ -i\Delta [A_i^a(\boldsymbol{x}) \Pi_i^a(\boldsymbol{x}) + \Pi_i^a(\boldsymbol{x}) A_i^a(\boldsymbol{x})] A_i^b(\boldsymbol{x}) \Pi_i^c(\boldsymbol{x}) \} = \exp\{ -ia [A_i^a(\boldsymbol{x})]^2 A_i^b(\boldsymbol{x}) \Pi_i^c(\boldsymbol{x})\} \nn\\
&\times \exp\{ -ib [\Pi_i^a(\boldsymbol{x})]^2 A_i^b(\boldsymbol{x}) \Pi_i^c(\boldsymbol{x})\} \exp\{ -ic [A_i^a(\boldsymbol{x})]^2 A_i^b(\boldsymbol{x}) \Pi_i^c(\boldsymbol{x})\} \exp\{ -id [\Pi_i^a(\boldsymbol{x})]^2 A_i^b(\boldsymbol{x}) \Pi_i^c(\boldsymbol{x})\} \,,
\eea
where $\Delta = -g^2 \sum_df^{dab}f^{dca} G(x_3,x_3)\Delta t /2$.
For $b\neq c$, the time evolution for each exponential on the right-hand side can be made diagonal in some basis and a product of evolutions driven by four Pauli $Z$ matrices. The basis transformation is done by applying local quantum Fourier transforms for a subset of colors. The implementation of the four-Pauli-$Z$ evolution is standard and involves six CNOT and one single-qubit rotation gates. If $b=c$, one needs to further decompose it using the same trick again and in the end obtains the product of $16$ exponentials, each of which has straightforward decompositions as above in a basis that can be achieved by local quantum Fourier transforms.

For the eight nonzero $\sum_a f^{abc}f^{ade}$ terms with $b=e$ but $c\neq d$, we have four exponentials, each involving $n_A^4$ tensor products of four Pauli $Z$ matrices for one spatial index. So we have $ 64 n_A^4 V$ such tensor products and thus need $384 n_A^4 V$ CNOT and $64 n_A^4 V$ single-qubit rotation gates, besides the costs for quantum Fourier transforms. Similarly for the eight nonzero $\sum_a f^{abc}f^{ade}$ terms with $c=d$ but $b\neq e$. For the 50 nonzero terms with $b=e$ and $c=d$, we have 16 exponentials and thus $1600$ evolution operators of four Pauli $Z$ matrices after taking into account the two spatial indices. For these $1600$ operators, we need $9600 n_A^4 V$ CNOT and $1600 n_A^4 V$ single-qubit rotation gates per Trotter step. 

When $x_3=x_3'$, there are also terms of the form
\bea
A_i^a(\boldsymbol{x}) \Pi_j^a(\boldsymbol{x}) A_j^b(\boldsymbol{x}) \Pi_i^c(\boldsymbol{x}) \,,
\eea
with $i\neq j$, $a\neq b$, and $a\neq c$. Their time evolution can be implemented similarly as the terms with $x_3\neq x_3'$ discussed above. The extra costs added by implementing these terms at most change $L-1$ to $L$ in the estimates below Eq.~\eqref{eqn:ffApiApi}.

\subsubsection{$H_{\Pi A \Pi}$}
The $H_{\Pi A \Pi}$ term in Eq.~\eqref{eqn:latticeHpiApi} does not involve the above subtlety: The operator
\bea
 f^{abc} \partial_i^{(L+)} \Pi_i^a({\boldsymbol x}) A_j^b({\boldsymbol x}')  \Pi_j^c({\boldsymbol x}') \,,
\eea
never contains non-commuting operators because of the antisymmetric $f^{abc}$. $H_{\Pi A \Pi}$ contains $432V^{4/3}$ nonzero terms, each of which can be decomposed into $n_A^3$ tensor products of three Pauli $Z$ matrices. So we need $1728n_A^3V^{4/3}$ CNOT and $432n_A^3V^{4/3}$ single-qubit rotation gates to implement one Trotter step, modulo costs for local quantum Fourier transforms. 

\subsection{Fermion Field Terms}
In order to maintain the anticommutation relation of the fermion fields as in Eq.~\eqref{eqn:anticommu} when mapping them onto qubits, one has to apply the Jordan-Wigner transformation or use the more efficient Bravyi-Kitaev encoding. Here we use the Jordan-Wigner transformation to be explicit. On a 3D spatial lattice, one can find a path going through every lattice site without repeating, via, e.g., a path with $x_1$ increasing first, then $x_2$, and finally $x_3$. This path defines a map
\bea
{\boldsymbol x}, f,i,\alpha \mapsto l_{f\alpha}^i({\boldsymbol x}) \,,
\eea
where $f,i,\alpha$ denote the fermion flavor ($f=1,2,\cdots,n_f$), fundamental color ($i=1,2,3$), and Dirac indices ($\alpha=1,2,3,4$), respectively. 
$l_{f\alpha}^i({\boldsymbol x})$ is an integer labeling the fermion's position and indices along the path. As the position and indices change, $l_{f\alpha}^i({\boldsymbol x})$ increases and the increment is always one. Specifically, at the $n$th spatial point [$n=x_1+L(x_2-1)+L^2(x_3-1)$ starts from 1] on the path, $l_{f\alpha}^i({\boldsymbol x})=12V(f-1)+12(n-1)+4(i-1)+\alpha$. With this map, we can implement an arbitrary two-fermion-field operator as
\bea
\label{eqn:JW}
\psi^{i\dagger}_{f\alpha}({\boldsymbol x}) \psi^j_{f'\beta}({\boldsymbol y}) \to \sigma^-_{l_{f\alpha}^i({\boldsymbol x})} \otimes \sigma^z_{l_{f\alpha}^i({\boldsymbol x})+1} \otimes \sigma^z_{l_{f\alpha}^i({\boldsymbol x})+2} \otimes \cdots \otimes \sigma^z_{l_{f'\beta}^j({\boldsymbol y})-1} \otimes \sigma^+_{l_{f'\beta}^j({\boldsymbol y})} \,,
\eea
where $\sigma^+$ and $\sigma^-$ are the raising and lowering Pauli matrices, respectively.
The number of Pauli $Z$ matrices in Eq.~\eqref{eqn:JW} is the implementation overhead due to the anticommutation relation of fermion fields.
For the QCD Hamiltonian, we always have $f=f'$. So one can actually apply Jordan-Wigner transformation for each flavor independently, thus reducing the overhead by removing the $12V(f-1)$ term in the expression of $l_{f\alpha}^i({\boldsymbol x})$.

\subsubsection{$H_f$}
The free fermion part $H_f$ in the Hamiltonian as shown in Eq.~\eqref{eqn:latticeHf} consists of three parts: the kinetic, mass, and Wilson terms. The kinetic term 
\bea
\psi_{f\alpha}^{i\dagger}(\boldsymbol{x}) (\gamma^0\gamma^j)_{\alpha\beta} [ \psi_{f\beta}^i(\boldsymbol{x} + \hat{j}) - \psi_{f\beta}^i(\boldsymbol{x} - \hat{j})]\,,
\eea
has an implementation overhead due to the different positions and Dirac indices. The difference in the Dirac indices gives overhead of at most two Pauli $Z$ matrices, e.g., when $\alpha=1$ and $\beta=4$. The relevant position differences here are $\pm \hat{j}$. Due to the Jordan-Wigner path, one $j$ gives $O(1)$ overhead, another gives $O(L)$ and the last one gives $O(L^2)$, which are at most 11, $12L-1$, and $12L^2-1$ Pauli $Z$ matrices, respectively. Adding these two types of overhead costs and accounting for the Pauli $\sigma^\pm$ in Eq.~\eqref{eqn:JW}, we find that the worst scenario corresponds to Pauli strings of length $15$, $12L+3$, and $12L^2+3$, respectively, to implement one kinetic term. Across the whole lattice, we multiply the cost by the volume $V$, the number of flavors $n_f$, the number of fundamental colors, which is three for QCD, a factor of two for $\pm \hat{j}$, and a factor of four for the four nonvanishing entries of $\gamma^0\gamma^j$. Putting things together, we find the cost to implement one-step time evolution of the fermion kinetic term is $24n_fV(36+24L+24L^2)$ CNOT and $72n_fV$ single-qubit rotation gates.

For the mass term
\bea
\psi_{f\alpha}^{i\dagger}(\boldsymbol{x}) \gamma^0_{\alpha\beta} \psi_{f\beta}^i(\boldsymbol{x})\,,
\eea
the only overhead may arise from the differences in the Dirac indices, which is at most two as mentioned above, corresponding to a Pauli string of length four to implement the mass term. Accounting for the whole lattice, flavors, colors, and nonzero entries of $\gamma^0$, we find the cost to implement the fermion mass term is $72 n_f V$ CNOT and $12n_fV$ single-qubit rotation gates.

Finally, for the Wilson term
\begin{align}
    \sum_{j=1,2,3} \psi^{i\dagger}_{f\alpha}({\boldsymbol x}) \gamma^0_{\alpha\beta} [\psi_{f\beta}^i({\boldsymbol x}+\hat{j}) + \psi_{f\beta}^i({\boldsymbol x}-\hat{j}) - 2\psi_{f\beta}^i({\boldsymbol x})] \,,
\end{align}
we note that the first two terms inside the square bracket are of a similar form to the kinetic term while the last term is of the same form as the mass term. So the last Wilson term can be implemented together with the mass, by adjusting the prefactor, adding no extra cost. The first two Wilson terms have different Dirac structures from the kinetic term [$\gamma^0_{\alpha\beta}$ vs. $(\gamma^0\gamma^j)_{\alpha\beta}$], so they cannot be implemented together. The costs for implementing the first two Wilson terms are the same as implementing the kinetic term. 

\subsubsection{$H_I$}
The $H_I$ term in Eq.~\eqref{eqn:latticeHI} involves
\bea
\label{eqn:AJ_one_term}
A_k^a({\boldsymbol x}) \psi_{f\alpha}^{i\dagger}({\boldsymbol x}) (\gamma^0\gamma^k)_{\alpha\beta} (T^a)^{ij} \psi_{f\beta}^j({\boldsymbol x}) \,.
\eea
The fermion part, which is at the same position, takes overhead of at most $10$ Pauli $Z$ matrices due to the differences in the color and Dirac indices. Using the Pauli decomposition of $A_k^a$ in Eq.~\eqref{eqn:Adecomp} and the Jordan-Wigner transformation in Eq.~\eqref{eqn:JW}, we see that one such term as in Eq.~\eqref{eqn:AJ_one_term} can be written as a sum of $n_A$ Pauli strings of length at most $13$, each of which requires at most $24$ CNOT and one single-qubit rotation gates to implement one Trotter step in time evolution. Summing over the whole lattice and all nonzero indices [two spatial indices $k=1,2$, four nonzero entries of $\gamma^0\gamma^k$, and 17 nonzero entries of $(T^a)^{ij}$ among all $a,i,j$] we find the cost to be $3264 n_f n_A V$ CNOT and $136 n_f n_A V$ single-qubit rotation gates per Trotter step.

\subsubsection{$H_{\Pi f}$}
The $H_{\Pi f}$ term in Eq.~\eqref{eqn:latticeHpif} contains the fermion color charge density operator
\bea
\psi^{i\dagger}_{f\alpha}({\boldsymbol x}) (T^a)^{ij} \psi_{f\alpha}^j({\boldsymbol x})\,,
\eea
in which the two fermion fields have the same position and Dirac index but different color indices. The Jordan-Wigner overhead is at most seven Pauli $Z$ matrices and one such charge density operator corresponds to a Pauli string of length at most nine. The $\partial_i^{(L+)} \Pi_i^a({\boldsymbol x}) = \Pi_i^a({\boldsymbol x}+\hat{i}) - \Pi_i^a({\boldsymbol x})$ term is diagonal in the local canonical conjugate variable basis and can be decomposed into a sum of single Pauli $Z$ matrices as used earlier. So $\Pi_k^a({\boldsymbol x}') \psi^{i\dagger}_{f\alpha}({\boldsymbol x}) (T^a)^{ij} \psi_{f\alpha}^j({\boldsymbol x})$ consists of Pauli strings of length at most 10, each requiring $18$ CNOT and one single-qubit rotation gates to implement one Trotter step. The number of nonzero $(T^a)^{ij}$ terms is 17 among all $a,i,j$. The number of Dirac indices is four. The total cost to implement one Trotter step for $H_{\Pi f}$ is $4896 n_f n_AV^{4/3}$ CNOT and $272 n_f n_AV^{4/3}$ single-qubit rotation gates.

\subsubsection{$H_{A\Pi f}$}
The $H_{A\Pi f}$ term in Eq.~\eqref{eqn:latticeHApif} can be similarly analyzed as $H_{\Pi f}$ except for an extra gauge field operator
\bea
f^{abc}A_k^b(\boldsymbol{x}) \Pi_k^c(\boldsymbol{x}) \psi^{i\dagger}_{f\alpha}({\boldsymbol x}') (T^a)^{ij} \psi_{f\alpha}^j({\boldsymbol x}') \,.
\eea
We emphasize again that due to the antisymmetric $f^{abc}$, the $A_k^b$ operator and $\Pi_k^c$ operator have different colors and thus commute. One can quantum Fourier transform the local field basis for one adjoint color but not for another to put them both in diagonal bases, in which each of them has a simple decomposition into a sum of Pauli $Z$ matrices. So $H_{A\Pi f}$ contains Pauli strings of length at most 11, each requiring $20$ CNOT and one single-qubit rotation gates to implement one Trotter step. The total cost per step is then $16000 n_f n_A^2V^{4/3}$ CNOT and $800 n_f n_A^2 V^{4/3}$ single-qubit rotation gates, since there are 100 nonzero entries in $\sum_af^{abc} (T^a)^{ij}$ among all $b,c,i,j$.

\subsubsection{$H_{ff}$}
Finally, the $H_{ff}$ term in Eq.~\eqref{eqn:latticeHff} couples two fermion color charge density operators
\bea
\psi^{i\dagger}_{f\alpha}({\boldsymbol x}) (T^a)^{ij} \psi_{f\alpha}^j({\boldsymbol x}) \psi^{k\dagger}_{f'\beta}({\boldsymbol x}') (T^a)^{k\ell} \psi_{f'\beta}^\ell({\boldsymbol x}')\,,
\eea
which contains Pauli strings of length at most 18, requiring $34$ CNOT and one single-qubit rotation gates to implement one Trotter step. The total cost for one step is $8160 n_f^2 V^{4/3}$ CNOT and $240 n_f^2 V^{4/3}$ single-qubit rotation gates, since there are 15 nonzero entries in $\sum_a (T^a)^{ij} (T^a)^{k\ell} $ among all $i,j,k,\ell$.

\subsection{Summary}
\label{sec:gate cost}
The costs of CNOT and single-qubit rotation gates to implement one Trotter step for each part of the axial gauge QCD Hamiltonian are summarized in Table~\ref{tab:costs}, as well as the costs for quantum Fourier transforms across the whole lattice. Parametrically, the dominant CNOT costs originate from the gauge interaction term $H_{A\Pi A\Pi}$ and the fermion kinetic term due to the Jordan-Wigner overhead, which scale as $O(n_A^4V^{4/3})$ and $O(V^{5/3})$, respectively, for fixed $n_f\leq 6$. The dominant cost of the single-qubit rotation gate comes from the $H_{A\Pi A\Pi}$ term, which scales as $O(n_A^4V^{4/3})$.

\begin{table}[h]
    \centering
    \begin{tabular}{|c|c|c|}
    \hline
    Components & CNOT &  Single-qubit rotation \\ \hline
    $H_A$  & $256 n_A^2 V + 432 n_A^3V + 2184 n_A^4V$  & $ 128 n_A^2 V + 108 n_A^3V + 364 n_A^4V$ \\ \hline
    $H_\Pi$  & $32 n_A^2 V$ & $16 n_A^2 V$ \\ \hline
    $H_{\Pi\Pi}$  & $160 n_A^2 V^{4/3}$ & $80 n_A^2 V^{4/3}$ \\ \hline
    $H_{A\Pi A\Pi}$  & $8736 n_A^4V^{4/3} + 10368 n_A^4V$ & $1456 n_A^4V^{4/3} + 1728 n_A^4V$ \\ \hline
    $H_{\Pi A\Pi}$  & $1728n_A^3 V^{4/3}$ & $432n_A^3 V^{4/3}$\\ \hline
    $H_f$  & $576n_fV(3+2V^{1/3}+2V^{2/3}) + 72n_fV$ & $156n_fV$ \\ \hline
    $H_I$  & $3264 n_fn_AV$ & $ 136 n_fn_AV$ \\ \hline
    $H_{\Pi f}$  & $4896 n_fn_AV^{4/3}$ & $272 n_fn_AV^{4/3}$ \\ \hline
    $H_{A\Pi f}$  & $16000 n_fn_A^2V^{4/3}$ & $800 n_fn_A^2V^{4/3}$ \\ \hline
    $H_{ff}$  & $8160 n_f^2 V^{4/3}$ & $240 n_f^2 V^{4/3}$ \\ \hline
    QFT & $16(n_A^2 - n_A + 3\lfloor n_A/2 \rfloor)V$ & $8(n_A^2+n_A)V$ \\ \hline
    \end{tabular}
    \caption{Number of CNOT and single-qubit rotation gates needed to implement the time evolution of each part of the Hamiltonian per Trotter step, as well as that for the local quantum Fourier transform across the whole lattice.}
    \label{tab:costs}
\end{table}

\begin{table}[thp]
    \centering
    \begin{tabular}{|c|c|}
    \hline
    $||[H_\Pi,H_A]||$  &  $O(A_{\rm max}\Pi_{\rm max}V) + O(gA^2_{\rm max}\Pi_{\rm max}V) + O(g^2A^3_{\rm max}\Pi_{\rm max}V)$ \\ \hline
    $||[H_\Pi,H_I]||$  &  $O(g\Pi_{\rm max}V)$ \\ \hline
    $||[H_\Pi,H_{A\Pi A\Pi}]||$  &  $O(g^2A_{\rm max}\Pi^3_{\rm max}V^{5/3})$ \\ \hline
    $||[H_\Pi,H_{\Pi A\Pi}]||$  &  $O(g\Pi^3_{\rm max}V^{5/3})$ \\ \hline
    $||[H_\Pi,H_{A\Pi f}]||$  &  $O(g^2\Pi^2_{\rm max}V^{5/3})$ \\ \hline
    $||[H_A,H_{\Pi\Pi}]||$ &  $O(A_{\rm max}\Pi_{\rm max}V^{5/3}) + O(gA^2_{\rm max}\Pi_{\rm max}V^{5/3}) + O(g^2A^3_{\rm max}\Pi_{\rm max}V^{5/3})$ \\ \hline
    $||[H_A,H_{A\Pi A\Pi}]||$  &  $O(g^2A^3_{\rm max}\Pi_{\rm max}V^{5/3}) + O(g^3A^4_{\rm max}\Pi_{\rm max}V^{5/3}) + O(g^4A^5_{\rm max}\Pi_{\rm max}V^{5/3})$ \\ \hline
    $||[H_A,H_{\Pi A\Pi}]||$  &  $O(gA^2_{\rm max}\Pi_{\rm max}V^{5/3}) + O(g^2A^3_{\rm max}\Pi_{\rm max}V^{5/3}) + O(g^3A^4_{\rm max}\Pi_{\rm max}V^{5/3})$ \\ \hline
    $||[H_A,H_{\Pi f}]||$  &  $O(gA_{\rm max}V^{5/3}) + O(g^2A^2_{\rm max}V^{5/3}) + O(g^3A^3_{\rm max}V^{5/3})$ \\ \hline
    $||[H_A,H_{A\Pi f}]||$  &  $O(g^2A^2_{\rm max}V^{5/3}) + O(g^3A^3_{\rm max}V^{5/3}) + O(g^4A^4_{\rm max}V^{5/3})$ \\ \hline
    $||[H_I,H_f]||$  &  $O(gA_{\rm max}V) + O(gmA_{\rm max}V) + O(grA_{\rm max}V)$ \\ \hline
    $||[H_I,H_{\Pi\Pi}]||$  &  $O(g\Pi_{\rm max}V^{5/3})$ \\ \hline
    $||[H_I,H_{A\Pi A\Pi}]||$  &  $O(g^3A_{\rm max}^2\Pi_{\rm max}V^{5/3})$ \\ \hline
    $||[H_I,H_{\Pi A\Pi}]||$  &  $O(g^2A_{\rm max}\Pi_{\rm max}V^{5/3})$ \\ \hline
    $||[H_I,H_{\Pi f}]||$  &  $O(g^2V^{5/3}) + O(g^2A_{\rm max}\Pi_{\rm max}V^{5/3})$ \\ \hline
    $||[H_I,H_{A\Pi f}]||$  &  $O(g^3A_{\rm max}V^{5/3}) + O(g^3A^2_{\rm max}\Pi_{\rm max}V^{5/3})$ \\ \hline
    $||[H_I,H_{ff}]||$  & $O(g^3A_{\rm max}V^{5/3})$  \\ \hline
    $||[H_f,H_{\Pi f}]||$  &  $O(g\Pi_{\rm max}V^{5/3}) + O(gm\Pi_{\rm max}V^{5/3}) + O(gr\Pi_{\rm max}V^{5/3}) $ \\ \hline
    $||[H_f,H_{A\Pi f}]||$  &  $O(g^2A_{\rm max}\Pi_{\rm max}V^{5/3}) + O(g^2m A_{\rm max}\Pi_{\rm max}V^{5/3}) + O(g^2r A_{\rm max}\Pi_{\rm max}V^{5/3}) $ \\ \hline
    $||[H_f,H_{ff}]||$  &  $O(g^2V^{5/3}) + O(g^2mV^{5/3}) + O(g^2rV^{5/3})$ \\ \hline
    $||[H_{\Pi\Pi},H_{A\Pi A\Pi}]||$  &  $O(g^2A_{\rm max}\Pi_{\rm max}^3 V^{7/3})$  \\ \hline
    $||[H_{\Pi\Pi},H_{\Pi A\Pi}]||$  &  $O(g\Pi_{\rm max}^3 V^{7/3})$ \\ \hline
    $||[H_{\Pi\Pi},H_{A\Pi f}]||$  &  $O(g^2\Pi_{\rm max}^2 V^{7/3})$ \\ \hline
    $||[H_{A\Pi A\Pi},H_{\Pi A\Pi}]||$  &  $O(g^3A^2_{\rm max}\Pi_{\rm max}^3 V^{7/3})$ \\ \hline
    $||[H_{A\Pi A\Pi},H_{\Pi f}]||$  &  $O(g^3A_{\rm max}\Pi_{\rm max}^2 V^{7/3})$ \\ \hline
    $||[H_{A\Pi A\Pi},H_{A\Pi f}]||$  &  $O(g^4A^2_{\rm max}\Pi_{\rm max}^2 V^{7/3})$  \\ \hline
    $||[H_{\Pi A\Pi},H_{\Pi f}]||$  &  $O(g^2\Pi_{\rm max}^2 V^{7/3})$  \\ \hline
    $||[H_{\Pi A\Pi},H_{A\Pi f}]||$  &  $O(g^3A_{\rm max}\Pi_{\rm max}^2 V^{7/3})$ \\ \hline
    $||[H_{\Pi f},H_{A\Pi f}]||$  &  $O(g^3\Pi_{\rm max} V^{7/3}) + O(g^3A_{\rm max}\Pi_{\rm max}^2 V^{7/3})$  \\ \hline
    $||[H_{\Pi f},H_{ff}]||$  &  $O(g^3\Pi_{\rm max} V^{7/3})$ \\ \hline
    $||[H_{A\Pi f},H_{ff}]||$  &  $O(g^4A_{\rm max}\Pi_{\rm max} V^{7/3})$ \\ \hline
    \end{tabular}
    \caption{Bounds on norms of commutators of different terms in the Hamiltonian. In the estimates, we have used $|G(x_3,x_3')|=O(L)$ for the Green's function and defined $m\equiv \sum_fm_f$.}
    \label{tab:norms}
\end{table}

Before closing this section, we make two remarks. First, the Trotter error can be bounded too. The number of Trotter steps needed to simulate the Hamiltonian evolution using first-order Trotterization up to a given time $t$ at a given accuracy $\epsilon$ is given by\footnote{The first-order formula is used here for concreteness. The error of a $2k$th-order Trotterization is estimated as~\cite{Childs:2019hts}
\begin{align}
    \epsilon \sim \sum_{i<j} || [H_i, H_j] || \frac{t^{2k+1} \eta^{2k-1}}{N_t^{2k}} \,,
\end{align}
where $\eta$ is the additional cost for each nested commutator in the higher-order Trotterization formula, due to the all-to-all coupling along the $x_3$ direction. A simple estimate based on the $H_{A\Pi A\Pi}$ term gives $\eta\sim g^2 A_{\rm max}\Pi_{\rm max} V^{2/3}$.}
\begin{align}
    N_t \sim \sum_{i<j} || [H_i, H_j] || \frac{t^2}{\epsilon} \,,
\end{align}
where $H_i$ and $H_j$ denote non-commuting terms in the Hamiltonian and $||\mathcal{O}||$ stands for the operator norm of $\mathcal{O}$. Table~\ref{tab:norms} lists bounds on the norms of different commutator terms. Further reduction of the Trotterization error may be achieved by using the techniques in Ref.~\cite{Tong:2021rfv}. Second, quantum Fourier transforms across the lattice only need to be applied a fixed number of times per Trotter step, some of which transform bases for all colors (for $H_\Pi$, $H_{\Pi\Pi}$, and $H_{\Pi f}$) and some only transform for a subset of colors (for $H_{A\Pi A\Pi}$, $H_{\Pi A\Pi}$, and $H_{A\Pi f}$). The sequence of these transforms can be optimized, which we leave to future practical implementation studies. We conclude that simulating real-time evolution up to a given time only requires polynomial quantum resources.

In practice, one may want to use an anisotropic lattice since the setup here treats the $x_3$ direction differently and the dominant cost comes from this direction. For a lattice of size $L_i$ along the $i$th direction, the costs can be obtained by replacing $V^{k/3}$ with $L_1$ for $k=1$, $L_1L_2$ for $k=2$, and $L_1L_2L_3^{k-2}$ for $k=3,4,5,7$ in Tables~\ref{tab:costs} and~\ref{tab:norms}.

\section{Conclusions}
\label{sec:conclusions}
In this paper, we studied quantum simulation of QCD by using the lattice Hamiltonian in axial gauge $A_3^a=0$ on an aperiodic lattice. The $A_0^a$ gauge field is analytically solved in terms of independent gauge and fermion fields and a lattice regulated Green's function of $(\partial_3)^2$, thus avoiding technical challenges associated with Gauss's law. The axial gauge condition $A_3^a=0$ is trivially maintained under (Trotterized) time evolution. The independent gauge field degrees of freedom are expressed in the local field basis, which can be efficiently transformed into the canonical conjugate variable basis via local quantum Fourier transforms. We proved a bound on the number of qubits needed to describe all states up to a given energy at a given accuracy. We then analyzed a time evolution algorithm that is based on Trotterization and Jordan-Wigner transformation for the fermion statistics. We showed that the numbers of CNOT and single-qubit rotation gates are bounded polynomially for simulating time evolution up to a given time at a given accuracy.

The Hamiltonian setup and the quantum algorithm studied here are generally applicable for SU($N_c$) non-Abelian gauge theories in more than one spatial dimension. One just needs to replace the SU(3) structure constant $f^{abc}$ with the corresponding one for SU($N_c$) and adjust the number of independent gauge fields according to the number of spatial dimensions. Qubit and gate counts can be bounded similarly as proved here. We plan to perform quantum simulation of SU(2) and SU(3) gauge theories in $2+1$ and then $3+1$ dimensions in future work, using the axial gauge setup. The algorithm developed here is just a starting point for simulating the axial gauge Hamiltonian. Further optimization of the algorithm will also be explored in future work.

\acknowledgments
This work is supported by the U.S. Department of Energy, Office of Science, Office of Nuclear Physics, InQubator for Quantum Simulation (IQuS)\footnote{\url{https://iqus.uw.edu/}} under Award Number DOE (NP) DE-SC0020970 via the program on Quantum Horizons\footnote{\url{https://science.osti.gov/np/Research/Quantum-Information-Science}}: QIS Research and Innovation for Nuclear Science.

\appendix
\section{Boundary Conditions with $A_i^a\neq 0$}
\label{app}
In the main text, we focus on the Dirichlet boundary condition in which all fields vanish just outside the cubic lattice, i.e., when $x_i=0$ or $x_i=L+1$ for $i=1,2,3$. For physical observables that do not involve gauge fields at spatial infinities this boundary condition is adequate, i.e., the zero mode does not contribute. For observables that depend on gauge fields $A_1^a$ and $A_2^a$ at spatial infinities, the axial gauge condition $A_3^a=0$ can still be applied and we just need to adjust the boundary condition such that these gauge fields just outside the lattice are set equal to some given values.
However, for observables that depend on the gauge field $A_3^a$ at spatial infinities, e.g., a straight spatial Wilson line connecting $x_3=- \infty$ and $x_3=\infty$, naively using the axial gauge condition $A_3^a=0$ does not give correct physical results as the zero momentum mode along $x_3$ contributes to the observable~\cite{Scheihing-Hitschfeld:2022xqx}. To cure the problem, we may set
\bea
A_3^a(\boldsymbol{x}) = A^a_\infty + \tilde{A}_3^a(\boldsymbol{x}) \,,
\eea
where $A_\infty^a$ denotes constants set to some given field values at spatial infinities. If there is more than one value of $A_3^a$ for a given $a$ at spatial infinities, we can repeat calculations for each value since the Hamiltonians associated with each value are decoupled under fixed boundary conditions that are aperiodic. This is just like adding a static charge when using the temporal gauge Hamiltonian to account for temporal Wilson lines going to infinite time or those that are periodic in Euclidean time~\cite{Pisarski:2022cuo}. The axial gauge condition now becomes $A_3^a=A_\infty^a$ or equivalently $\tilde{A}_3^a(\boldsymbol{x})=0$. Then we have $F_{03}^a = -\partial_3 A_0^a + gf^{abc} A_0^b A_\infty^c$ and solving $A_0^a$ now needs inverting $(D_3)^2 A_0^a = Q^a$, which is more involved but can be done as the nonzero $A_3^a$ field is constant. For observables that involve $A_3^a$ with one or two spatial coordinates at infinities, it would be much easier to rotate the coordinates such that $A_3^a$ can be set to zero at spatial infinities.

\bibliography{main}
\end{document}